\documentclass{kluwer}    
\usepackage{graphicx}

\newcommand{\gtappeq}{\raisebox{-0.6ex}{$\,\stackrel
{\raisebox{-.2ex}{$\textstyle >$}}{\sim}\,$}}
 
\newdisplay{guess}{Conjecture}

\begin{document}                                                                                   
\begin{article}
\begin{opening}         
\title{Is GRS 1915+105 a microquasar?} 
\runningtitle{Is GRS 1915+105 a microquasar?}
\author{Christian R. \surname{Kaiser}}  
\institute{School of Physics \& Astronomy, University of Southampton, Southampton, SO17 1BJ, UK}
\author{J.L. \surname{Sokoloski}}
\institute{Harvard-Smithsonian Center for Astrophysics, Cambridge, MA 02138, USA}
\author{Katherine F. \surname{Gunn}}
\institute{School of Physics \& Astronomy, University of Southampton, Southampton, SO17 1BJ, UK}
\author{Catherine \surname{Brocksopp}}
\institute{Mullard Space Science Laboratory, University College London, Dorking, Surrey, RH5 6NT, UK}
\runningauthor{Kaiser, Sokoloski, Gunn \& Brocksopp}
\date{October 1, 2004}

\begin{abstract}
The large mechanical luminosity of the jets of GRS 1915+105 should give rise to luminous emission regions, similar to those observed in radio galaxies, where the jets interact with the gas surrounding the source.  However, no radio synchrotron emission of the expected morphology has been found. Here we present the results of a study suggesting that radio bremsstrahlung from the compressed and heated ISM in front of the jets should be detectable, while the synchrotron lobes may be too faint. We identify these jet impact sites with two well-known IRAS regions. This identification suggests a distance of GRS 1915+105 of $6.5\pm1.6$\,kpc, significantly closer than the usually assumed distance of 11 to 12\,kpc. We discuss the implications of this reduced distance estimate. The apparent motion of small-scale jet components is not superluminal, so if superluminal motion is required for an object to be termed a microquasar, GRS 1915+105 actually does not qualify. The mass of the black hole in the system is increased to $21\pm9$\,M$_{\odot}$ while the mechanical luminosity of the jets is reduced to 14\% of the Eddington luminosity. 
\end{abstract}

\end{opening}           

\section{A primer in large-scale radio structures caused by jets}

Jets can efficiently carry an enormous amount of energy from the central engine to sites far removed from their acceleration regions. The dissipation of the jet's energy can give rise to spectacular large-scale structures, often observed in the radio. It is these structures we are trying to identify for the powerful jets of the microquasar GRS 1915+105. 

Jets are usually thought to be ballistic after their initial acceleration. Ballistic jets are not luminous as they do not dissipate much of their bulk kinetic energy. However, unless the gas surrounding the jet flow has a very steep negative density gradient, the jet will eventually have to collimate by developing a reconfinement shock \cite{sf91}. After passing through this shock, the jet is in pressure equilibrium with its surroundings and can become susceptible to turbulent disruption. If the jet flow becomes fully turbulent, its energy is dissipated slowly over a large volume and therefore the resulting radiation is often not very luminous. For radio galaxies, such a turbulent jet flow would result in a radio structure of type FR\,I \cite{fr74}. If the jet stays laminar, it will end in a strong shock where it impacts on the surrounding gas. This shock or working surface is identified as the very luminous radio hot-spots of radio galaxies with an FR\,II-type morphology. After passing through the hot-spot, the jet material inflates more diffuse, but still luminous radio lobes \cite{ps74}. The radio emission of the hot-spots and the lobes is synchrotron radiation from relativistic electrons accelerated by the shock at the end of the jet \cite{hr74}. Laminar jets deliver virtually all their energy to the working surface where it is dissipated in a small volume. Therefore they can give rise to much more luminous large-scale radio structures than their turbulent counterparts. It is not surprising that the original morphological distinction between FR\,Is and FR\,IIs is also reflected in a sharp division in radio luminosity between the two classes. This distinction demonstrates that the detection of the large-scale radio structure of any jet source will be much simplified if the jets of this source stay laminar throughout their length.

The expansion of the jets and their lobes drives a bow shock into their gaseous environment. The shock-compressed and heated gas will also emit radiation. The frequency and spectrum of this radiation depends crucially on the conditions in the gas. For the hot ($\gtappeq 10^7$\,K) and therefore fully ionised IGM surrounding the large-scale structures of radio galaxies, the shocked gas will emit thermal bremsstrahlung at X-ray frequencies (e.g. \citeauthor{swa02}, \citeyear{swa02}). The colder ISM surrounding microquasar jets may only be partially ionised by the passage of the bow shock, and features like recombination lines and emission from heated dust could arise. The partially ionised ISM can still produce thermal bremsstrahlung, but due to the lower temperatures this radiation will be emitted at radio wavelengths rather than in X-rays. Laminar jets typically dissipate about half of their total bulk kinetic energy in the shock-compressed gas surrounding their lobes \cite{ka96b} and so this material may contribute substantially to the overall emission from the large-scale structure caused by the jets.

In summary, we may expect to detect the following components of the large-scale structure caused by laminar microquasar jets:
\begin{itemize}
\item Compact radio synchrotron emission regions at the shocks at the end of the jets.
\item Extended, diffuse radio synchrotron emission lobes.
\item Extended radio bremsstrahlung, recombination lines and heated dust emission surrounding the lobes. 
\end{itemize}

The synchrotron radio lobes of microquasar jets have been detected in several sources (SS433 \citeauthor{dhgm98}, \citeyear{dhgm98}; 1E1740.7-2942 \citeauthor{mrc93}, \citeyear{mrc93}; Cir X-1 \citeauthor{sch93}, \citeyear{sch93}). However, in the case of GRS 1915+105 searches have been unsuccessful \cite{rm98}. In the following we will show that some signs of the interaction of the jets of GRS 1915+105 with the ISM have been observed already, but they do not include the radio synchrotron lobes.

\section{What can we expect in the case of GRS 1915+105?}

As we pointed out in the previous section, our chances for detecting the large-scale jet structure of GRS 1915+105 are greatly enhanced if its jets stay laminar after coming into pressure equilibrium with their surroundings. Following a simple momentum conservation argument \cite{cr91}, we show in \citeauthor{kgs04} \shortcite{kgs04} that the jets should stay laminar over a length of at least a few to several parsec even if they are in direct contact with a very dense ($n \sim 3000$\,cm$^{-3}$) ISM. For an average ISM density of $n\sim 1$\,cm$^{-3}$, this estimate increases by a factor of roughly 50. Also, it is likely that the lobes inflated by the jets extend all the way back to the acceleration region \cite{sh02}, thereby creating a protective environment and greatly extending the range over which the jet flows can remain laminar.

In \citeauthor{kgs04} \shortcite{kgs04} we demonstrate that the impact site of one of these laminar jets should generate a bremsstrahlung flux of 
\begin{equation}
F_{1\,{\rm GHz}} \sim 5 \left(\frac{Q}{10^{37}\,{\rm erg}\,{\rm s}^{-1}} \right)  \left( \frac{T}{10^4\,{\rm K}} \right)^{-2} \left( \frac{x}{0.01} \right) \left( \frac{D}{\rm kpc} \right)^{-2}\,{\rm Jy},
\end{equation}
where $Q$ is the total energy transport rate of the jet. Note that this flux does not depend on the density of the ISM nor on the volume occupied by the emitting gas. The jet impact sites should therefore be easily detectable in the radio at GHz frequencies. Furthermore, given the likely temperature of the compressed ISM of $T \sim 10^4$\,K, we expect to detect recombination lines from hydrogen. If the ISM in front of the jet is multi-phase, then we expect to detect emission lines from various molecules as well as dust emission. 

Estimating the radio synchrotron flux of the lobes is more complicated as this depends crucially on the physical size of the lobes and the gas density in their environment. We defer a discussion of this emission to the next section.

\section{Identification of the jet impact sites}

\begin{figure}
\centerline{
\includegraphics[width=8cm]{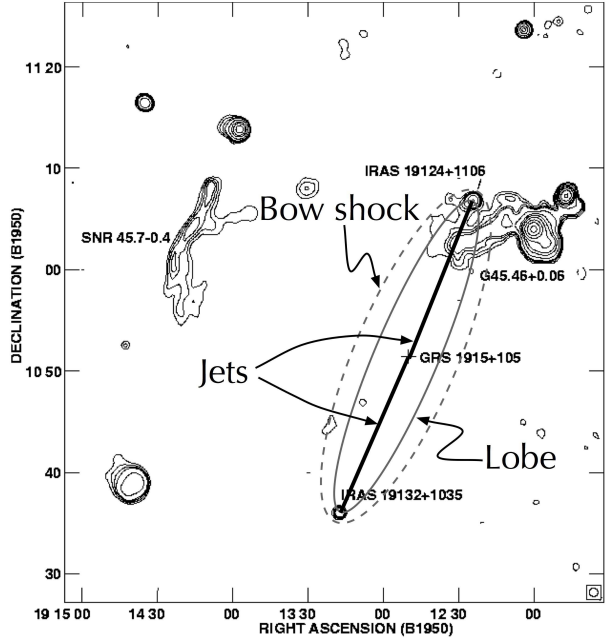}}
\caption{Radio map of the surroundings of GRS 1915+105 taken with the VLA at 20\,cm with the postulated large-scale jet structure superimposed. }
\label{environs}
\end{figure}

Given the expected properties of the jet impact sites, they can only be identified with IRAS 19124+1106 and IRAS 19132+1035 (see Figure \ref{environs} and \citeauthor{kgb04}, \citeyear{kgb04}). Their flat radio spectra \cite{rm98} and luminosities are consistent with radio bremsstrahlung. The detected radio recombination lines and molecular emission lines \cite{crm01} support the identification. The observed IR fluxes are consistent with emission from heated dust \cite{kgb04}. 

These emission properties are also consistent with an interpretation of the IRAS objects as star-forming H{\sc ii} regions. However, a chance alignment of two such H{\sc ii} regions along the jets of a microquasar is small \cite{kgs04}. Furthermore, the elongated, non-thermal emission feature connected to IRAS 19132+1035 noted by \citeauthor{rm98} \shortcite{rm98} points back to the position of GRS 1915+105 and may be a transient hot-spot at the end of the southern jet. We believe that these results lend strong support to our interpretation.

In \citeauthor{kgb04} \shortcite{kgb04} we present upper limits for the surface brightness of potential radio lobes employing the usual minimum energy arguments (e.g. \citeauthor{ml94}, \citeyear{ml94}). We find that these limits lie below current detection limits of radio maps of the vicinity of GRS 1915+105 \cite{rm98}. However, in that paper we neglected the fact that most of the ISM compressed by the bow shock in front of the jets will not be collisionally ionised at the temperature of $T \sim 1.2\times 10^4$\,K, inferred from the width of observed emission lines. Taking into account the appropriate ionisation fraction of about 2\% implies a density of the unshocked ISM in the vicinity of the jets of  $n\sim 3000$\,cm$^{-3}$. This new estimate also changes our upper limit for the surface brightness of the lobes due to radio synchrotron to
\begin{equation}
S_{1\,{\rm GHz}} \sim 100 \left( \frac{D}{\rm kpc} \right)^{1/8} \left( k+1 \right)^{-7/4}\,{\rm mJy\,beam}^{-1},
\end{equation}
where $D$ is the distance of GRS 1915+105 and $k$ is the ratio of energy in non-radiating particles to the sum of the energy stored in the relativistic electrons and the energy in the magnetic field. For ideal minimum energy conditions we require $k=0$. We assume a beam size of 4", appropriate for the observations of \citeauthor{rm98} \shortcite{rm98}. 

Clearly the synchrotron radio lobes of GRS 1915+105 should be detected in current radio maps which have a limit of 0.2\,mJy\,beam$^{-1}$, unless $k > 34$, which is not unreasonable (e.g. \citeauthor{ab78}, \citeyear{ab78}). Therefore, given the estimate of S$_{\rm 1 GHz}$ above, the non-detection of synchrotron lobes implies a jet composition other than a pure pair plasma.

\section{Implications of our identification}

If the two IRAS sources are indeed the impact sites of the jets of GRS 1915+105, then this microquasar must be located at the same distance of $6.5\pm1.6$\,kpc \cite{kgb04} as opposed to the traditionally assumed distance of 11 to 12\,kpc. This revised distance is consistent with all previous distance estimates \cite{fgmmpssw99,dgr00,dmr00,cc04} except one \cite{gcm01} and has some profound implications for the inferred properties of this system. 

The jets of GRS 1915+105 travel with a bulk velocity of $0.66\pm0.15$\,c at an angle of $52.6\pm7.2^{\circ}$ to our line of sight. The apparent velocity of the jet ejecta on the sky is $0.66\pm0.18$\,c, which is clearly not superluminal. The luminosity of any unbeamed emission will be reduced by a factor $\sim 3$ compared to the usually assumed distance. The lower limit on the power of the jets during the formation of the moving ejecta observed on small angular scales is reduced by more than two orders of magnitude to $2 \times 10^{37}$\,erg\,s$^{-1}$. The total, time-averaged mechanical luminosity of the jets derived from the dynamical model of the jet expansion is $Q \sim 4 \times 10^{38}$\,erg\,s$^{-1}$. Note that this estimate is a factor 50 larger than our original estimate \cite{kgb04} because of the revised density estimate for the ISM in the vicinity of GRS 1915+105 from taking into account the partial ionisation of the hydrogen gas \cite{kgs04}. Finally, the mass of the black hole in the system is increased to $21\pm9$\,M$_{\odot}$, resulting in a total mechanical jet luminosity of 14\% of the Eddington luminosity.

The question of whether or not GRS 1915+105 is a microquasar hinges mainly on whether the definition of the term `microquasar' includes the requirement for superluminal jets. If it does, then the first object with apparently superluminal jets in the Galaxy is not a microquasar, and it is therefore somewhat different from other microquasars that clearly show highly relativistic jet flows (e.g. Cir X-1, \citeauthor{fwj04}, \citeyear{fwj04}). It is also interesting to note that GRS 1915+105 appears to be a binary system with a far less energetic jet, but a more massive black hole than previously appreciated.

\end{article}

\begin{thebibliography}{}

\bibitem[\protect\citeauthoryear{Bell}{1978}]{ab78}
Bell, A.~R.: 1978, `The acceleration of cosmic rays in shock fronts -- II'.
\newblock {\em MNRAS} {\bf 182}, 443.

\bibitem[\protect\citeauthoryear{Cant\'o and Raga}{1991}]{cr91}
Cant\'o, J. and A.~C. Raga: 1991, `Mixing layers in stellar outflows'.
\newblock {\em ApJ} {\bf 372}, 646.

\bibitem[\protect\citeauthoryear{Chapius and Corbel}{2004}]{cc04}
Chapius, C. and S. Corbel: 2004, `On the optical extinction and distance of GRS
  1915+105'.
\newblock {\em A\&A} {\bf 414}, 659.

\bibitem[\protect\citeauthoryear{Chaty et~al.}{2001}]{crm01}
Chaty, S., L.~F. Rodr{\'\i}guez, I.~F. Mirabel, T.~R. Geballe, Y. Fuchs, A.
  Claret, C.~J. Cesarsky, and D. Cesarsky: 2001, `A search for the possible
  interactions between ejections from GRS 1915+105 and the surrounding
  interstellar medium'.
\newblock {\em A\&A} {\bf 366}, 1035.

\bibitem[\protect\citeauthoryear{Dhawan et~al.}{2000a}]{dgr00}
Dhawan, V., W.~M. Goss, and L.~F. Rodr{\'\i}guez: 2000a, `Small-scale structure
  in Galactic HI absorption towards GRS 1915+105'.
\newblock {\em ApJ} {\bf 540}, 863.

\bibitem[\protect\citeauthoryear{Dhawan et~al.}{2000b}]{dmr00}
Dhawan, V., I.~F. Mirabel, and L.~F. Rodr{\'\i}guez: 2000b, `AU-scale
  synchrotron jets and superluminal ejecta in GRS 1915+105'.
\newblock {\em ApJ} {\bf 543}, 373.

\bibitem[\protect\citeauthoryear{Dubner et~al.}{1998}]{dhgm98}
Dubner, G.~M., M. Holdaway, W.~M. Goss, and I.~F. Mirabel: 1998, `A
  High-Resolution Radio Study of the W50-SS 433 System and the Surrounding
  Medium'.
\newblock {\em AJ} {\bf 116}, 1842.

\bibitem[\protect\citeauthoryear{Falle}{1991}]{sf91}
Falle, S. A. E.~G.: 1991, `Self--similar jets'.
\newblock {\em MNRAS} {\bf 250}, 581.

\bibitem[\protect\citeauthoryear{Fanaroff and Riley}{1974}]{fr74}
Fanaroff, B.~L. and J.~M. Riley: 1974, `The morphology of extragalactic radio
  sources of high and low luminosity'.
\newblock {\em MNRAS} {\bf 167}, 31.

\bibitem[\protect\citeauthoryear{Fender et~al.}{2004}]{fwj04}
Fender, R., K. Wu, H. Johnston, T. Tzioumis, P. Jonker, R. Spencer, and M. {van
  der Klis}: 2004, `An ultra-relativistic outflow from a neutron star accreting
  gas from a companion'.
\newblock {\em Nat.} {\bf 427}, 222.

\bibitem[\protect\citeauthoryear{Fender et~al.}{1999}]{fgmmpssw99}
Fender, R.~P., S.~T. Garrington, D.~J. McKay, T.~W.~B. Muxlow, G.~G. Pooley,
  R.~E. Spencer, A.~M. Stirling, and E.~B. Waltman: 1999, `MERLIN observations
  of the relativistic ejections from GRS 1915+105'.
\newblock {\em MNRAS} {\bf 304}, 865.

\bibitem[\protect\citeauthoryear{Greiner et~al.}{2001}]{gcm01}
Greiner, J., J.~G. Cuby, and J. McCaughrean: 2001, `An unusually massive
  stellar black hole in the Galaxy'.
\newblock {\em Nat.} {\bf 414}, 522.

\bibitem[\protect\citeauthoryear{Hargrave and Ryle}{1974}]{hr74}
Hargrave, P.~J. and M. Ryle: 1974, `Observations of Cygnus A with the 5km radio
  telescope'.
\newblock {\em MNRAS} {\bf 166}, 305.

\bibitem[\protect\citeauthoryear{Heinz}{2002}]{sh02}
Heinz, S.: 2002, `Radio lobe dynamics and the environment of microquasars'.
\newblock {\em A\&A} {\bf 388}, L40.

\bibitem[\protect\citeauthoryear{Kaiser and Alexander}{1997}]{ka96b}
Kaiser, C.~R. and P. Alexander: 1997, `A self--similar model for extragalactic
  radio sources'.
\newblock {\em MNRAS} {\bf 286}, 215.

\bibitem[\protect\citeauthoryear{Kaiser et~al.}{2004a}]{kgb04}
Kaiser, C.~R., K.~F. Gunn, C. Brocksopp, and J.~L. Sokoloski: 2004a, `Revision
  of the properties of the GRS 1915+105 jets: Clues from the large-scale
  structure'.
\newblock {\em ApJ} {\bf 612}, 332.

\bibitem[\protect\citeauthoryear{Kaiser et~al.}{2004b}]{kgs04}
Kaiser, C.~R., K.~F. Gunn, J.~L. Sokoloski, and C. Brocksopp: 2004b, `Is the
  microquasar GRS 1915+105 really superluminal?'.
\newblock {\em ApJ Letters: submitted}.

\bibitem[\protect\citeauthoryear{Longair}{1994}]{ml94}
Longair, M.~S.: 1994, {\em High energy astrophysics}.
\newblock Cambridge University Press.

\bibitem[\protect\citeauthoryear{Mirabel et~al.}{1993}]{mrc93}
Mirabel, I.~F., L.~F. Rodr{\'\i}guez, B. Cordier, J. Paul, and F. Lebrun: 1993,
  `VLA observations of the hard X-ray sources 1E1740.7-2942 and GRS1758-258'.
\newblock {\em A\&A Supp.} {\bf 97}, 193.

\bibitem[\protect\citeauthoryear{Rodr{\'\i}guez and Mirabel}{1998}]{rm98}
Rodr{\'\i}guez, L.~F. and I.~F. Mirabel: 1998, `The surroundings of the
  superluminal source GRS 1915+105'.
\newblock {\em A\&A} {\bf 340}, L47.

\bibitem[\protect\citeauthoryear{Scheuer}{1974}]{ps74}
Scheuer, P. A.~G.: 1974, `Models for extragalactic radio sources with a
  continous energy supply from a central object'.
\newblock {\em MNRAS} {\bf 166}, 513.

\bibitem[\protect\citeauthoryear{Smith et~al.}{2002}]{swa02}
Smith, D.~A., A.~S. Wilson, K.~A. Arnaud, Y. Terashima, and A.~J. Young: 2002,
  `A Chandra X-Ray Study of Cygnus A. III. The Cluster of Galaxies'.
\newblock {\em ApJ} {\bf 565}, 195.

\bibitem[\protect\citeauthoryear{Stewart et~al.}{1993}]{sch93}
Stewart, R.~T., J.~L. Caswell, R.~F. Haynes, and G.~J. Nelson: 1993, `Circinius
  X-1: a runaway binary with curved radio jets'.
\newblock {\em MNRAS} {\bf 261}, 593.

\end{thebibliography}
\end{document}